\documentclass[12pt]{article}
\usepackage[english]{babel}
\usepackage{amsmath}
\usepackage{amssymb}
\usepackage{mathtools}
\usepackage{amsthm}
\usepackage{nicefrac}
\usepackage{pifont} 
\usepackage{microtype}
\usepackage{caption}
\usepackage{xcolor}
\usepackage{subcaption}
\usepackage{nccmath}
\usepackage{soul}
\usepackage{appendix}
\usepackage{graphicx}
\usepackage{natbib}
\usepackage{tikz}
\usepackage{xcolor}
\usepackage{hyperref}
\PassOptionsToPackage{linktocpage}{hyperref}
\hypersetup{
           breaklinks=true,   
           colorlinks=true,   
           linkcolor = blue,
            urlcolor  = blue,
            citecolor = blue,
            anchorcolor = blue,
        }
\usetikzlibrary{decorations.pathreplacing}
\usepackage{algorithm}
\usepackage[noend]{algpseudocode}
\algnewcommand{\Input}{\textbf{Input:} }
\algnewcommand{\Output}{\textbf{Output:} }
\algnewcommand{\Draw}{\textbf{draw} }
\usepackage{url}
\usepackage{setspace}
\newcommand{\blind}{0}
\usepackage{changes}
\usepackage{todonotes}

\begin{document}

\def\spacingset#1{\renewcommand{\baselinestretch}%
{#1}\small\normalsize} \spacingset{1}

\if0\blind
{
  \title{\bf A Point Mass Proposal Method for Bayesian State-Space Model Fitting}
  \author{Mary Llewellyn, Ruth King, V\'{i}ctor Elvira and Gordon Ross \\
  \normalsize{\it{School of Mathematics and Maxwell Institute for Mathematical Sciences}}, \\ \normalsize{\it{University of Edinburgh, Edinburgh, UK}}}
  \date{}
  \maketitle
} 

\if1\blind
{
  \bigskip
  \bigskip
  \bigskip
  \begin{center}
    {\LARGE\bf A Point Mass Proposal Method for Bayesian State-Space Model Fitting}
\end{center}
  \medskip
} \fi

\begin{abstract}
State-space models (SSMs) are commonly used to model time series data where the observations depend on an unobserved latent process. However, inference on the model parameters of an SSM can be challenging, especially when the likelihood of the data given the parameters is not available in closed-form. One approach is to jointly sample the latent states and model parameters via Markov chain Monte Carlo (MCMC) and/or sequential Monte Carlo approximation. These methods can be inefficient, mixing poorly when there are many highly correlated latent states or parameters, or when there is a high rate of sample impoverishment in the sequential Monte Carlo approximations. We propose a novel block proposal distribution for Metropolis-within-Gibbs sampling on the joint latent state and parameter space. The proposal distribution is informed by a deterministic hidden Markov model (HMM), defined such that the usual theoretical guarantees of MCMC algorithms apply. We discuss how the HMMs are constructed, the generality of the approach arising from the tuning parameters, and how these tuning parameters can be chosen efficiently in practice. We demonstrate that the proposed algorithm using HMM approximations provides an efficient alternative method for fitting state-space models, even for those that exhibit near-chaotic behavior.
\end{abstract}

\noindent%

\spacingset{1.1}

\section{Introduction}\label{intro}

State-space models (SSMs) describe time series observations as dependent on an unobserved system process \citep{Durbin_2012}. The unobserved system process of an SSM consists of a sequence of continuously-valued latent states that evolve through time but are not observed directly. Instead, observations are modeled as a function of these unobserved latent states via the observation process. SSMs have been applied to various problems, including the modeling of inflation rates \mbox{\citep{Koopman_2004}}, neuron responses \citep{Smith_2003, Lin_2019}, animal movement \citep{Patterson_2008, Mcclintock_2012, AugerMethe_2016}, and population dynamics \citep{Buckland_2004, King_2009, King_2012, Newman_2014}.

Fitting an SSM usually refers to inference on the model parameters given the observations. However, outside of the discrete-state hidden Markov model (HMM) case \citep{Rabiner_1989}, or special continuous cases, for example, linear Gaussian SSMs \citep{Kalman_1960, Durbin_2012}, a closed-form expression for the likelihood does not exist, leading to model fitting challenges. The various approaches to approximating the likelihood include the use of linear Gaussian approximations \citep{Julier_1997, Wan_2000}, numerical methods using Laplace approximations or the HMM approximations of point mass filters \citep{Bucy_1971, Koyama_2010, Langrock_2013, Thygesen_2017, Herliansyah22}, the use of sequential Monte Carlo to obtain an unbiased approximation to the likelihood \citep{Beaumont_2003, Andrieu_2009, Kantas_2015, Deligiannidis_2018}, and Bayesian data augmentation approaches \citep{Fruhwirth_2004,Fearnhead_2011,Borowska23}. A review of these methods, and others, is given in \linebreak \cite{Newman_2022}.

Here, we focus on Bayesian data augmentation approaches, which can be applied to general SSMs by specifying the latent states as additional auxiliary variables that need to be estimated. In doing so, the data augmentation approach forms the joint posterior distribution of the model parameters and additional latent states \citep{Tanner_1987, Hobert_2011}. The resulting joint likelihood of the parameters and states is typically available in closed-form, permitting the application of standard MCMC algorithms, such as the Gibbs sampler, Metropolis-Hastings algorithms, or Metropolis-within-Gibbs algorithms. An estimate of the marginal posterior distribution of the model parameters can then be obtained by simply retaining the sampled parameter values and disregarding the latent state samples to effectively marginalize over the states. It is, however, typically challenging to efficiently sample the latent state vector within a Metropolis-Hastings algorithm, since there are typically many highly correlated latent states, leading to poor mixing \citep{Fruhwirth_2004}. Updating the states one at a time typically performs very poorly due to the high correlation between states \citep{Fearnhead_2011}, whilst designing good block-update proposal distributions can be challenging \citep{Shephard_1997}.

Historically, data augmentation has been combined with numerical approximation, sequential Monte Carlo, and/or MCMC to fit the parameters of an SSM. When an SSM is well-approximated by Gaussian distributions, efficient proposal distributions for the latent states in a data augmentation approach include those derived from Laplace approximation \citep{Kristensen_2016} and extensions to the Kalman filter \citep{Giordani_2010}. In the general case, more computationally intensive methods are usually required, such as the particle Gibbs (PG) sampler \citep{Lindsten_2012, Chopin_2015, Fearnhead_2016} or particle Gibbs with ancestor sampling (PGAS) \citep{Lindsten_2014}. PG and PGAS are classes of particle MCMC method \citep{Andrieu_2010} which use sequential Monte Carlo to update all states simultaneously conditional on the parameters. In their use of sequential Monte Carlo, PG and PGAS methods rely on few assumptions about the underlying state and observation distributions. However, these methods can be computationally expensive, particularly if many samples are required in the sequential Monte Carlo steps due to sample impoverishment \citep{Rainforth_2016, Wang_2017}. One approach to overcoming sample impoverishment is the introduction of deterministic resampling, which uses a deterministic grid over the state space to retain samples that may otherwise be discarded in the sequential Monte Carlo steps \citep{Li_2012}. Although we do not develop a particle MCMC approach here, the use of a deterministic grid to efficiently sample the latent states is related to our approach.

In this paper, we propose a novel and efficient approach to Metropolis-within-Gibbs data augmentation. In particular, we focus on the challenging aspect of efficiently updating the latent states conditional on the model parameters and present the \textit{point mass proposal Metropolis-Hastings} (PMPMH) algorithm, proposing candidate values for (blocked) latent state values in two steps. The first step uses a discretization of the state space into a set of pre-defined intervals (a grid) and deterministic approximations to reduce the SSM to an HMM. We then sample a sequence of intervals from the HMM-approximated distribution of the latent states given the model parameters, employing the forward-filtering backward-sampling (FFBS) algorithm \citep{Rabiner_1989}. In the second step of the algorithm, we sample values from within the intervals using a continuous proposal distribution. Once the candidate values have been proposed using the two-step procedure, we correct for the discretization error imposed by the approximate HMM using a Metropolis-Hastings acceptance step. As with standard data augmentation algorithms, the model parameters are then updated conditionally on the latent states using standard Gibbs or Metropolis-within-Gibbs updates. Further to describing a general method for combining HMM approximations and MCMC steps, we discuss, in detail, several tuning parameters that can be used to improve the performance of the algorithm. Such tuning parameters include the block size for the latent states (i.e. the number of latent states to update simultaneously), the number of intervals used in the HMM approximation, and the distribution of the intervals in the state space. 

The first step of the proposed algorithm is related to embedded HMMs and point mass filtering, which both use the tractable discrete case of an SSM, an HMM, to sample latent states. Embedded HMMs \citep{Neal_2003, Shestopaloff_2013, Finke_2016, Shestopaloff_2018} construct an HMM in the indices of stochastically-generated `pool’ states, and sample states such that they target the correct posterior distribution. Here, we use the embedding of an HMM via deterministic grid cells to ensure latent states are proposed with reasonable posterior mass. Point mass filters, similarly to the proposed approach, use a deterministic grid to reduce the SSM to an HMM \citep{Bucy_1971, Kitagawa_1987, Langrock_2012, Langrock_2013}. The HMM can be used to approximate the likelihood or posterior distribution of the model parameters directly. As such, point mass filters can be (relatively) computationally inexpensive to implement, but generally produce biased estimators. Further, the bias and variance of estimators often depend on the grid (for example, the number and location of the grid cells and their coverage of the state space), and the integration method used to approximate the HMM \citep{deValphine_2002, Matousek_2019}. Instead, we use the grid cells simply to inform a Metropolis-Hastings proposal distribution and are thus able to define the HMM and grid cells coarsely (and inexpensively), while being able to achieve convergence to the correct posterior distribution, correcting for the bias introduced by the discretization.

\pagebreak
The rest of the paper is structured as follows. We start by defining SSMs and HMMs and model fitting procedures for SSMs via Bayesian data augmentation in Section \ref{background}. We describe the proposed data augmentation scheme in Section \ref{method}, the PMPMH algorithm, in which we use a grid-based HMM approximation within a proposal distribution for the latent states. In Section \ref{practical_considerations}, we provide three particular cases of the algorithm, each given by a different method for defining the grid cells, and describe when each can be usefully applied. Finally, we illustrate our approach with two case studies in Section \ref{applications}, including a near-chaotic system, where our proposed approach demonstrates a substantial improvement in performance compared to alternative approaches considered. We conclude in Section \ref{discussion} with a discussion of our proposed algorithm and further possible avenues of research.

\section{State-space model inference}\label{background}
Here we define SSMs, the notation used throughout the paper, and the special case of discrete-valued system processes: HMMs. We then discuss the challenges of fitting SSMs to data and computational approaches in Section \ref{background:parameter_inference}, motivating our proposed data augmentation algorithm. 

We consider time series data observed at discrete (regular) time points up to time $T$, $y_{1:T}=(y_1,\dots,y_T)$. An SSM models these data via (\emph{i}) a system process of unobserved (latent) states,  $x_{1:T}=(x_1,\dots,x_T)$; and (\emph{ii}) an observation process linking the latent process with the observed $y_{1:T}$. We allow the states to be continuous, taking values in some set $x_t \in \chi$ for each $t$, and describe their evolution using initial state and transition distributions. Similarly, the observation process is described using an observed state distribution. As standard, we assume a first-order Markov process for the system process and that the observed data are conditionally independent given the underlying latent states, that is, the observation process is only a function of the latent state at time $t$.

Given the set of model parameters, collectively denoted $\pmb{\theta}$, the SSM can be described mathematically by 

\begin{align*}
\mbox{\emph{Initial state distribution}:} \qquad   p(x_1 & \vert \pmb{\theta}), \\
\mbox{\emph{State transition distribution}:} \qquad      p(x_t &\vert x_{t-1}, \pmb{\theta}), \hspace{2mm} t=2, \dots, T, \\ 
\mbox{\emph{Observed state distribution}:} \qquad  p(y_t &\vert x_t, \pmb{\theta}), \hspace{2mm} t=1, \dots, T.
\end{align*} 

A particular case of an SSM is an HMM, in which the state space is discrete and finite. In other words, the states at time $t$, $x_t$, take values in the set $\{1, \dots, N\}$. Therefore, when the SSM considered is an HMM, we simply re-define the initial state and transition probabilities for $k, n=1, \dots, N$ as

\begin{align*}
    \mbox{\emph{Initial state probabilities}:}& \qquad P(x_1=n \vert \pmb{\theta}), \hspace{2mm} \\
    \mbox{\emph{State transition probabilities}:}& \qquad P(x_t=n \vert x_{t-1}=k, \pmb{\theta}), \hspace{2mm} t=2, \dots, T.
\end{align*}

\subsection{Model fitting}\label{background:parameter_inference}
We assume that we are primarily interested in inference on the model parameters, $\pmb{\theta}$, although this may also extend to the latent states, $x_{1:T}$, depending on the application. In general, inference on the posterior distribution for $\pmb{\theta}$ requires a closed-form expression for the observed data (or marginal) likelihood, $p(y_{1:T} \vert \pmb{\theta})$. The joint likelihood of the model parameters and latent states is given by 

\begin{equation}
    p(x_{1:T}, y_{1:T} \vert \pmb{\theta}) = p(x_1 \vert \pmb{\theta}) \prod_{t=2}^T p(x_t \vert x_{t-1}, \pmb{\theta}) \prod_{t=1}^T p(y_t \vert x_t, \pmb{\theta}).\label{eq:joint_likelihood}
\end{equation}

A closed-form expression for the marginal likelihood typically only exists in the linear Gaussian case, where $p(y_{1:T} \vert \pmb{\theta})$ can be calculated using the Kalman filter \citep{Kalman_1960, Durbin_2012}, or when the state space is discrete and finite (an HMM), where the marginalization of Equation (\ref{eq:joint_likelihood}) amounts to summation over the state space \citep{Rabiner_1989}. In general, however, the marginal likelihood is intractable for continuous SSMs, and standard MCMC implementations cannot be used. One solution is to use pseudo-marginal likelihood methods \citep{Beaumont_2003, Andrieu_2009}, replacing the marginal likelihood with an unbiased estimate to formulate a valid Metropolis-Hastings approach, but we do not focus on these here. 

Instead, we consider a Bayesian data augmentation approach. In this case, we form the joint posterior distribution over the model parameters and latent states \citep{Tanner_1987, Hobert_2011}, $p(x_{1:T}, \pmb{\theta} \vert y_{1:T}) \propto p(\pmb{\theta}) p(x_{1:T}, y_{1:T} \vert \pmb{\theta})$, thus using the joint likelihood in Equation (\ref{eq:joint_likelihood}) directly. Since, in general, a closed-form expression exists for the joint likelihood term up to proportionality, the proportional likelihood can be evaluated, and standard MCMC approaches can be used targeting $p(x_{1:T}, \pmb{\theta} \vert y_{1:T})$. Samples from the (marginal) posterior distribution of the parameters, $p(\pmb{\theta} \vert y_{1:T})$, are then obtained by disregarding the state samples and retaining the parameter samples.

To sample from $p(x_{1:T}, \pmb{\theta} \vert y_{1:T})$, a common approach is to sample the latent states and parameters in turn via their conditional distributions, simplifying the parameter updates \citep{Carter_1994, Fruhwirth_2004, Durbin_2012}. A Metropolis-within-Gibbs algorithm can be applied to target the joint distribution: in each iteration, we update the latent states conditional on the model parameters; and the model parameters conditional on the latent states (for which we now have a closed conditional likelihood expression). However, designing an efficient proposal distribution for the underlying state vector can be challenging. Ideally, we would be able to design a proposal distribution to update all states simultaneously and minimize the correlation between consecutive samples. In general, this state distribution is intractable, and designing an efficient approximation to the state distribution can be challenging since it is often high-dimensional in the number of temporal states. One approach updates all states simultaneously using a particle Gibbs (PG) algorithm \citep{Chopin_2015, Murphy_2016, Lindsten_2014}; values for the entire state vector are proposed at each iteration from a conditional sequential Monte Carlo algorithm. In this approach, proposed values for the entire state vector are always accepted and samples from the correct posterior distribution can be obtained. However, PG methods can suffer from poor mixing sample impoverishment: degenerating particles being disregarded upon resampling, impacting mixing and convergence \citep{Wang_2017, Rainforth_2016}. Instead of updating all states simultaneously, for example via PG updates, an alternative approach is to use a single-update algorithm: updating each latent state individually to reduce the required dimension of the proposal distribution, but this approach typically leads to very poor mixing due to correlation in the state process \citep{Fearnhead_2011, King_2011}. 

A compromise between single and global updates of the latent states is to use block updates, simultaneously updating $\ell$ consecutive states \citep{Shephard_1997}. The posterior conditional distribution of the states in a given block is, generally, of non-standard form, thus it can be challenging to define an efficient proposal distribution that accounts for the correlation between the states. We propose the use of the discrete analogy to SSMs, HMMs, to define an informative and efficient proposal distribution for blocks of latent states.

\section{Point mass proposal Metropolis-Hastings}\label{method}

In this section, we propose the \textit{point mass proposal Metropolis-Hastings} (PMPMH) algorithm. The setting is a Metropolis-within-Gibbs algorithm for data augmentation. The model parameters, $\pmb{\theta}$, are updated conditionally on $x_{1:T}$ using a standard Metropolis-Hastings or Gibbs update, and we propose the PMPMH algorithm to sample Metropolis-Hastings proposals for the latent states given the model parameters.
 
Step 1 of the algorithm initially converts the SSM into an HMM by discretizing the state space at each time point, forming a deterministic grid, and the HMM is approximated deterministically. Grid cells are then sampled from the associated (discretized and approximate) posterior distribution, conditional on the model parameters, using the standard forward-filtering backward-sampling (FFBS) algorithm for HMMs \citep{Rabiner_1989} \linebreak at a computational cost of $\mathcal{O}(NT)$ for $N$ grid cells at each time point. Given the sampled grid cells, Step 2 proposes values for $x_{1:T}$ from within the cells using some specified (bounded) continuous proposal distribution. We then correct for the approximate sampling distribution using a Metropolis-Hastings step targeting the correct posterior distribution.

For pedagogical purposes, we make a few assumptions in our initial description of the algorithm, for example, we derive the algorithm for one-dimensional state spaces. Extensions to higher-dimensional state spaces are possible, and we demonstrate the algorithm in the two-dimensional case with an example in Section \ref{applications}, and discuss scalability further in Section \ref{discussion}. However, we note here that there may be practically increased computational demands for higher-dimensional state spaces. For notational simplicity, we present the algorithm for SSMs where the state space and the number of grid cells are the same for each time point, that is, the state space at all time points is denoted by $\chi$. \linebreak \pagebreak However, the generalization to state spaces that vary over time is immediate by adapting the notation. Finally, we initially describe the algorithm in terms of updating the full set of latent states, with a version of the algorithm for block updates of the states described later.

\subsection{Step 1: sampling a grid cell trajectory }\label{method:grid_cell_traj}
In this initial step, we formulate a discrete HMM representation of the SSM, approximate the HMM, and sample a trajectory from the discrete approximation.

\subsubsection{Discretization to an HMM}
\begin{figure*}[h]
\centering 
\includegraphics[width=0.8286\paperwidth]{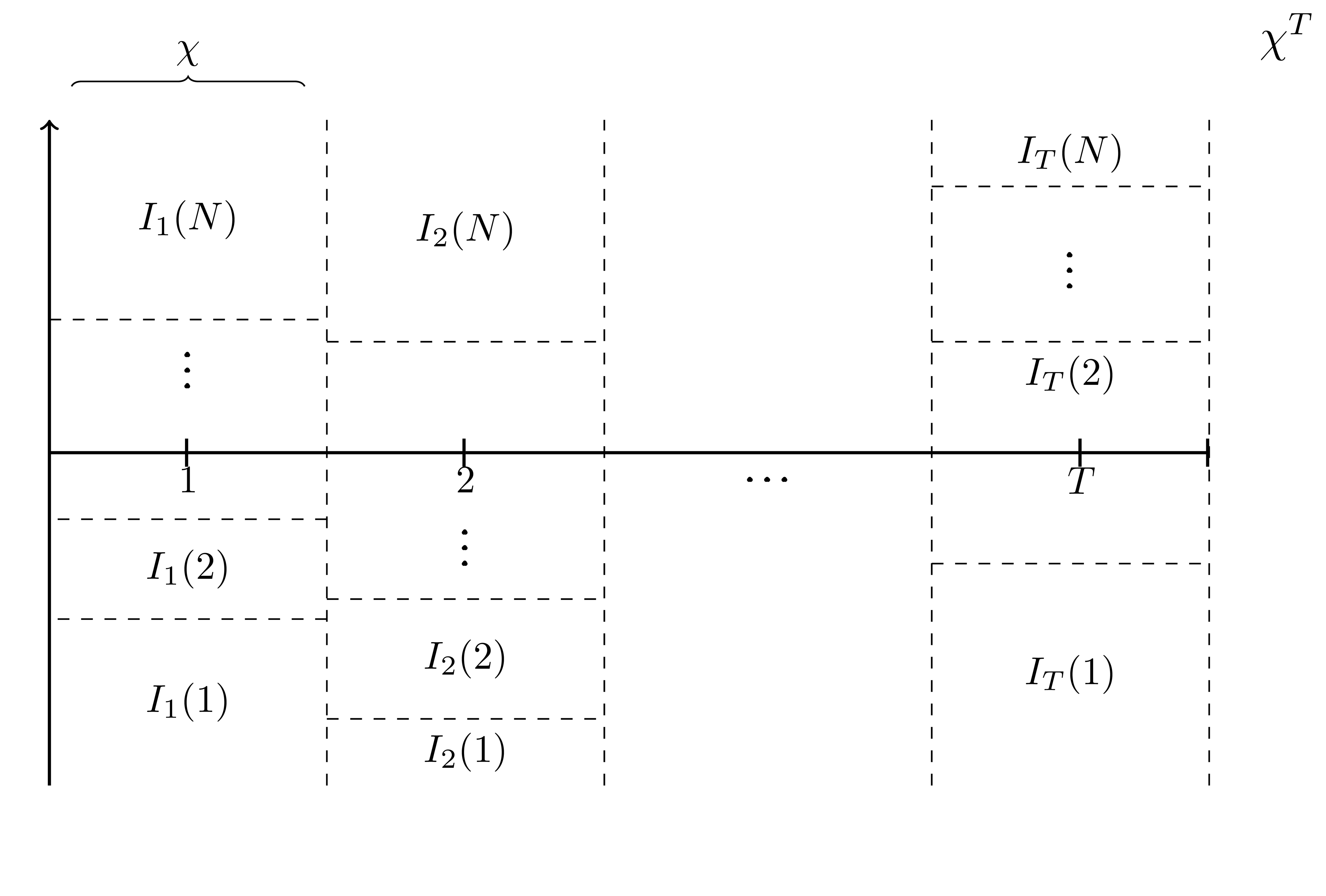}
\caption{Example partition of a state space $\chi^T$ for each time $t$. Each grid cell is labeled by the interval it covers}\label{fig:partition}
\end{figure*}

First, we propose a partition of the whole state space into a grid. The state space at each time $t$, $\chi$, is partitioned into intervals, forming grid cells. The $N$ grid cells at time $t$ are denoted $I_t(n),$ $n=1, \dots, N$ and span $\chi$ with no overlap. The respective outer grid cells have infinite length if the state space is unbounded. An example of such a partition is shown in Figure \ref{fig:partition}.

\newpage 
The grid cells discretize the state space. Thus, the grid cell indices are the discrete states of an HMM. For the grid cells given by $\{I_t(1), \dots I_t(N)\}$, the indices are simply $\{1, \dots N\}$, with dynamics given by the SSM density of each grid cell. Mathematically, let $B_t$ denote the random variable of the grid cell indices under the specified grid cell boundaries so that $B_t \in \{1,\dots,N\}$ for $t=1,\dots,T$. Then the HMM probabilities are given for states $k, n \in \{1, \dots, N\}$ as follows: 

\begin{align*}
       \mbox{\emph{Initial state probabilities}:}& \qquad P(B_1=n \vert \pmb{\theta}), \nonumber\\
    \mbox{\emph{State transition probabilities}:}& \qquad P(B_t = n \vert B_{t-1}=k, \pmb{\theta}), \hspace{2mm} t=2, \dots, T, \nonumber \\ 
     \mbox{\emph{Observed state distribution}:}& \qquad p(y_t \vert B_t=n, \pmb{\theta}), \hspace{2mm} t=1, \dots, T. 
\end{align*}

These HMM probabilities under the original SSM are therefore defined by integrals of the state space equations over the given intervals: 
\begin{align}
    P(B_1=n \vert \pmb{\theta}) &= \int_{I_1(n)} p(x_1 \vert \pmb{\theta}) dx_1, \nonumber \\
    P(B_t = n \vert B_{t-1}=k, \pmb{\theta}) &= \int_{I_t(n)} \int_{I_{t-1}(k)} p(x_t \vert x_{t-1}, \pmb{\theta}) dx_{t-1} dx_t, \hspace{2mm} t=2, \dots, T, \nonumber \\
     p(y_t \vert B_t=n, \pmb{\theta}) &= \int_{I_t(n)} p(y_t \vert x_t, \pmb{\theta}) dx_t, \hspace{2mm} t=1, \dots, T.\label{eq:ints}
\end{align}

\subsubsection{Approximate HMM}

In general, the probabilities associated with the HMM, given in Equation (\ref{eq:ints}), are not available in closed form, thus we approximate these integrals. To formulate an approximation to the HMM under the considered SSM, we use simple (fast) deterministic methods, for example, the midpoint rule or other Riemann sum methods. Mathematically, we define $L_t(n)$ as the length of the interval $I_t(n)$, with the length of any infinite grid cells set at some arbitrary (finite) length (discussed further in Appendix \ref{other_practicals}). Within each grid cell, we choose a set of node points, $\{\xi_t^s(n)\}_{s=1}^S$. The set of nodes is defined \emph{a priori} and is used to approximate the density functions across each grid cell by a polynomial. In the simple case where $S=1$ (that is, one node in each grid cell such as the mid-point), we have that $\{\xi_t^s(n)\}_{s=1}^S=\xi_t(n)$. The approximate HMM, for grid cells indices $k, n=1, \dots, N$, can thus be given by the mid-point approximations:

\begin{align*}
    \hat{P}(B_1 = n \vert \pmb{\theta}) &\propto L_1(n) p(\xi_1(n) \vert \pmb{\theta}), \\ 
    \hat{P}(B_t=n \vert B_{t-1}=k, \pmb{\theta}) &\propto L_t(n)  L_{t-1}(k)  p(\xi_t(n) \vert \xi_{t-1}(k), \pmb{\theta}), \hspace{2mm} t=2, \dots, T, \\ 
    \hat{P}(y_t \vert B_t=n, \pmb{\theta}) &\propto L_t(n) p(y_t \vert \xi_t(n), \pmb{\theta}), \hspace{2mm} t=1, \dots, T,
\end{align*}
\newpage 
Each approximate probability is appropriately normalized so that they sum to one and are thus valid probability mass functions. Once the probability mass function approximations have been obtained, grid cell indices are proposed from

\begin{equation*}
    \hat{P}(B_{1:T} \vert y_{1:T}, \pmb{\theta}) \propto \hat{P}(B_1 \vert \pmb{\theta})\hat{P}(y_1 \vert B_1, \pmb{\theta}) \prod_{t=2}^T \hat{P}(B_t \vert B_{t-1}, \pmb{\theta})\hat{P}(y_t \vert B_t, \pmb{\theta}).
\end{equation*}

We sample indices from this distribution, denoted $b'_{1:T}$, using the forward-filtering backward-sampling (FFBS) algorithm in \cite{Rabiner_1989}.

\subsection{Step 2: sampling a point trajectory}\label{method:point_traj}

Step 2 proposes continuous values for the states which will be subsequently corrected in Metropolis-Hastings steps. Given the discrete grid cell indices sampled previously, we propose values for $x_{1:T}$ conditional on the corresponding intervals $\{I_1(b'_1),\dots,I_T(b'_T)\}$. This amounts to sampling $x_{1:T}$ from the domain $I_1(b'_1)\times I_2(b'_2)\times \cdots \times I_T(b'_T)$. For simplicity, we consider proposal distributions for each $x_t$ independently and independently of $\pmb{\theta}$, giving proposal distributions of the form $q(x_t \vert x_t \in I_t(b'_t))=q(x_t \vert B_t=b'_t)$, for $t=1,\dots,T$. For example, a natural choice is to propose each $x_t$ from a uniform proposal distribution over $I_t(b'_t)$ for finite $I_t(b'_t)$, or a truncated Gaussian distribution if $I_t(b'_t)$ has infinite length. However, any distribution with domain $I_t(b'_t)$ is applicable.

\subsection{The PMPMH proposal distribution}
The process of sampling a grid cell trajectory and sampling a value for $x_{1:T}$ defines an independent proposal distribution on $\chi^T$. The density function of the proposed trajectories is therefore the probability of selecting the grid cell indices from a specified set of grid cell boundaries, combined with the density of the (continuous) values within those grid cells. Let $B_{1:T}$ denote the random variables of the grid cell indices under the proposed grid, which may change across iterations with intervals $I_t(n), \hspace{2mm} n=1, \dots, N, \hspace{2mm} t=1, \dots, T$. Then the density of the subsequently proposed state values, $x'_{1:T}$, is given by

\begin{equation}
        q(x'_{1:T} \vert y_{1:T}, \pmb{\theta}) = \hat{P}(B_{1:T}=b'_{1:T} \vert y_{1:T}, \pmb{\theta}) \prod_{t=1}^T q(x'_t \vert B_t=b'_t). 
    \label{eq:PMPMH_prop}
\end{equation}

The resulting proposed trajectory, $x'_{1:T}$, is retained according to the Metropolis-Hastings acceptance probability, given by 
\begin{equation}
    p(x_{1:T}, x'_{1:T} \vert \pmb{\theta}) = \min \left( 1, \frac{ p(x'_{1:T} \vert y_{1:T}, \pmb{\theta}) q(x_{1:T} \vert y_{1:T}, \pmb{\theta}) }{p(x_{1:T} \vert y_{1:T}, \pmb{\theta})q(x'_{1:T} \vert y_{1:T}, \pmb{\theta})} \right),\label{pacc}
\end{equation}
where $p(x_{1:T} \vert y_{1:T}, \pmb{\theta})$ denotes the posterior conditional distribution of $x_{1:T}$ given $y_{1:T}$ and $\pmb{\theta}$, $q(x'_{1:T} \vert y_{1:T}, \pmb{\theta})$ is as in Equation (\ref{eq:PMPMH_prop}), and $q(x_{1:T} \vert y_{1:T}, \pmb{\theta})$ uses the same definition for the grid cells, that is, their indices, $b_{1:T}$, are such that $x_t \in I(b_t)$ for all $t$. 

The PMPMH method for obtaining a sample of $x_{1:T}$ conditional on $\pmb{\theta}$ is summarized in \linebreak Algorithm \ref{PMPMH_alg}. For completeness, we also include the updates of $\pmb{\theta}$ conditional on $x_{1:T}$, which can be updated using standard Metropolis-Hastings or Gibbs updates conditional on the current chain value of the states.

\begin{algorithm*}[h]
\caption{PMPMH algorithm}\label{PMPMH_alg}
\textbf{Input:} initial values $x_{1:T}^{(0)}$, $\pmb{\theta}^{(0)}$ and a number of iterations $M$. A Gibbs or Metropolis-Hastings sampling scheme to update $\pmb{\theta}$ from $p(\pmb{\theta} \vert x_{1:T}, y_{1:T})$. 

\hfill

\textbf{for} iterations $m=1, \dots M$ \textbf{do}
\begin{algorithmic}[1]
\State{Update $\pmb{\theta}^{(m)}$ from $p(\pmb{\theta} \vert x_{1:T}^{(m-1)}, y_{1:T})$}
\Statex

\State{Define a grid with indices $B_{1:T}$}
\For{$t=1,..,T$} \label{fwd} \Comment{(Step 1)}
\For{$k=1,...,N$} 
\State{calculate $\hat{P}(B_t = n \vert y_{1:t}, \pmb{\theta}^{(m)})$} 
\EndFor
\EndFor
\State{sample $b'_T$ from $\{n, \hat{P}(B_T=n \vert y_{1:T}, \pmb{\theta}^{(m)})\}_{n=1}^N$}

\For{$t=T-1,...,1$} 
\State{sample $b'_t$ from $\{n, \hat{P}(B_t=n \vert B_{t+1}=b'_{t+1}, y_{1:t}, \pmb{\theta}^{(m)})\}_{n=1}^N$}\label{bwd}

\EndFor
\Statex

\For{$t=1,...,T$} \label{point_loop} \Comment{(Step 2)}
\State{sample $x'_t$ from  $q(x_t \vert B_t=b'_t)$} \label{point_sample}
\EndFor
\Statex

\State{sample $u \sim U(0,1)$} \label{acc_start}

\If{$u<p(x_{1:T}^{(m-1)}, x'_{1:T} \vert \pmb{\theta})$} \Comment{Equation (\ref{pacc})}
\State{$x_{1:T}^{(m)} \leftarrow x'_{1:T}$}
\Else{
$x_{1:T}^{(m)} \leftarrow x^{(m-1)}_{1:T}$} \label{acc_end}
\EndIf

\end{algorithmic}
\end{algorithm*}

\subsection{Block updates of the latent states}\label{method:blocking} 

Updating all of the states simultaneously may be inefficient if there are many latent states. We can, instead, update the states in smaller blocks: for a block of $\ell$ states starting at time $t$, we sample $x_{t:t+\ell-1}$ from a proposal distribution of the form $q(x_{t:t+\ell-1} \vert x_{t-1}, x_{t+\ell}, y_{t:t+\ell-1}, \pmb{\theta})$. 

A block PMPMH proposal distribution is given by a simple adaptation of the case where all states are updated simultaneously, conditioning on the current states at either side of the block; $x_{t-1}$ and $x_{t+\ell}$. Therefore, in Step 1, we condition on the current grid cell indices, $b_{t-1}$ and $b_{t+\ell}$, such that $x_{t-1} \in I(b_{t-1})$ and $x_{t+\ell} \in I(b_{t+\ell})$ under the current grid definition. We then use FFBS to sample proposed indices from the grid cells in the block. In Step 2, we simply define the proposal distribution, as before, for each latent state in the given block, $q(x_s \vert B_s=b'_s)$, for $s=t, \dots, t+\ell-1$. The PMPMH algorithm for block updates is summarized in Appendix \ref{code:blocking}. 

We note that it is possible to parallelize the updates of the latent states to improve computational efficiency. For example, blocks of states can be updated in parallel provided that the states to be updated and those being conditioned on do not overlap. For the sake of comparison with other methods, we do not parallelize our computation but simply note its possibility. See also \cite{King_2023} for further discussion on paralellization within a state-space model fitting context.

\subsection{Conditions for convergence}
A benefit of the proposed method is that estimators of integrable functions are consistent under some mild conditions ensuring the validity of the proposal distribution in the Metropolis-Hastings steps.

\mbox{\textbf{Theorem}} \textit{The PMPMH algorithm provides consistent ergodic average estimators of integrable functions with respect to $p(x_{1:T}, \pmb{\theta} \vert y_{1:T})$ if,}
\textit{
\begin{itemize}
    \item [1. ] for all $b_t \in \{1, \dots, N\}$, $t \in \{1, \dots, T\}$, the HMM probabilities are defined such that $\hat{P}(B_{1:T} = b_{1:T} \vert y_{1:T}, \pmb{\theta})>0$ and the proposal distributions within each grid cell are defined such that $q(x_t\vert B_t=b_t)>0$ for all $x_t \in I_t(b_t)$, and 
    \item [2. ] 
    \begin{itemize}
        \item[ (a)] the grid cells are defined and fixed at the start of the algorithm or
        \item[ (b)] the grid cells are defined as a function of the current MCMC iteration parameter or state values.
    \end{itemize} 
\end{itemize}}

\begin{proof}
We assume that the transition kernel on the parameter space is constructed such that it admits $p(\pmb{\theta}\vert x_{1:T}, y_{1:T})$ as its limiting distribution. By standard results from Metropolis-within-Gibbs algorithms \citep{Latuszynski_2013, Gamerman_2006}, proving that the PMPMH algorithm provides consistent estimators of integrable functions reduces to ensuring that the transition kernel on the blocks of $\ell$ states converge to the correct conditional posterior distribution. We therefore show that irreducibility and detailed balance are satisfied by the proposal distributions with respect to the correct conditional posterior distribution for each block of $\ell$ states.

By construction, we assume that $\cup_{n=1}^N I_t(n) = \chi$ for all $t=1, \dots, T$. Condition 1 implies that \mbox{$q(x_{1:T} \vert y_{1:T}, \pmb{\theta})>0$} for all $x_{1:T} \in \chi^T$, thus irreducibility is satisfied since any value in the state space can be proposed at each iteration of the algorithm. Note that Condition 1 can be ensured by arbitrarily thresholding the HMM probabilities above zero and attributing non-zero density within each grid cell using many standard distributions.

It is immediate that detailed balance holds under Condition 2(a) that the grid cells are defined and fixed at the start of the Metropolis-Hastings algorithm. It is also straightforward to show that detailed balance holds under Condition 2(b) that the grid cells are defined as a function of the current MCMC iteration parameter or state values.

Thus, Condition 1 and Condition 2(a) or 2(b) ensure that irreducibility and detailed balance are satisfied by the PMPMH proposal distribution, and the algorithm provides consistent ergodic average estimators of integrable functions with respect to \mbox{$p(x_{1:T}, \pmb{\theta} \vert y_{1:T})$.}
\end{proof}

\section{Defining the grid cells}\label{practical_considerations}
We have, so far, given a broad framework for using an approximate HMM as a Metropolis-Hastings proposal distribution. The method fundamentally relies on the use of a deterministic grid to discretize the state space. There are, however, many different ways to define the grid cells (their size and location), possibly resulting in substantially different proposal distributions and computational costs. We see in Section \ref{applications} that the efficiency of the algorithm is highly dependent on the choice of grid cell definition and resulting proposal distribution. Since the optimal choice depends on the application, we provide the reader with three approaches for defining the grid cells and describe when each can be usefully (and efficiently) applied. We discuss the other practical decisions in Appendix \ref{other_practicals}.

\subsection{Approach 1: equal grid cells}
The first approach sets all finite cells equally sized and the same across all time points, that is, $I_t(n) = I_s(n)$ for all $t,s = 1, \dots, T$ and $n=2, \dots, N-1$ (assuming an infinite lower and upper bound on the latent states). We assume that these equally sized grid cells are centered around the mean of the data, $\frac{1}{T} \sum_{t=1}^T y_t$, spanning a range of the state space denoted by $\mathcal{S}$, though this approach could be adapted by centering the grid cells around another function of the data. 

If the model parameters do not vary with time, only one transition matrix needs to be calculated per MCMC iteration, or when the model parameters are updated. For a fixed number of grid cells, this approach has the lowest computational cost of the approaches considered. However, since the grid cells are the same for each time point, but the regions of high posterior density may change at each time point, the grid cells should cover a large range of the state space to ensure coverage of these high-density regions and good mixing. This may require many grid cells and a greater computational cost if the high posterior regions of the SSM are highly non-uniform over time, and hence this approach is most efficient for SSMs with uniform such regions. 

\subsection{Approach 2: data-driven quantiles}
If the high posterior density regions vary over time, a time-inhomogeneous approximation to the HMM transition matrix may be more efficient than the previous approach. Hence, we set grid cells in this approach using the observed data at each time point. In particular, in the implementations of this paper, we set the grid cell boundaries at the quantiles of one-dimensional Gaussian distributions centered around each observed data point. For each time point, $t$, we set the boundaries at the quantiles of $X$ where $X \sim N(y_t, {\sigma_y}^2)$. The variance of the Gaussian distribution, ${\sigma_y}^2$, is set at a scalar value, or as a function of the current observation process variance within the MCMC iterations and can be used to control the range that the majority of the grid cells cover via pilot tuning. The resulting quantiles are simply rounded if integer values are required. Note that simple extensions to this approach may include using a different distribution to set the quantiles, provided that its domain is in the state space.

When the model parameters are updated, a transition matrix needs to be recalculated for each time point. It is, however, possible to reduce the computational cost of this approach by using the same transition matrix across several time points, centering, for example, on a single data point or the mean of the corresponding sub-series. By using the observed data at each time point, we aim to approximately concentrate the grid cells in areas that are likely to have high posterior mass at each time point. This method may be most efficient if the observed data used in the centering is a good proxy for the underlying state process and its dependencies.

\subsection{Approach 3: latent state quantiles}
The previous approaches define the grid cells independently of the state process, which may result in slow mixing if the latent states of the SSM are highly correlated over time. The accuracy of the proposal distribution and mixing of the latent states can be improved through the definition of the grid cells. \cite{Matousek_2019} show empirically that the accuracy of deterministic grid cell approximations to the posterior distribution can be improved by placing the grid cells according to rough approximations to the posterior distribution. Thus, here we set the grid cells using the current latent states in the MCMC iterations at each time $t$. The underlying intuition is that, after the MCMC algorithm has converged, the current states reflect a sample from the posterior distribution. Thus, over several iterations, grid cells centered around the current state lead to proposed values that are distributed similarly to the (conditional) posterior distribution \citep{Haario_1999}, improving the accuracy of the proposal distribution and mixing of the MCMC steps.

To formulate grid cells using the current state, we let ${x_t}^{(m-1)}$ denote the current state in the MCMC iterations at time $t$, then similarly to the previous approach, we define the grid cell boundaries at the quantiles of a Gaussian distribution $X$ for each $t$, where $X \sim N({x_t}^{(m-1)}, {\sigma_x}^2)$. As with the previous approach, the variance of the Gaussian distribution used to define the grid cells, ${\sigma_x}^2$, controls the span of the finite cells. Again, this variance could be set, for example, as a fixed value (chosen via pilot-tuning), as a function of the current estimate of the system process variance, or as a function of the current state within the MCMC iterations. We note that setting the majority of the grid cells to cover a small range relative to the high posterior density ranges worked well in practice (see \linebreak Section \ref{applications}).

In this approach, the transition matrices need to be recalculated at each time point for every change in the model parameters but will typically require fewer grid cells than the other approaches to achieve good mixing. \\ 

In general, there is a trade-off between the extent to which the grid cells are \emph{well-placed} and the associated computational expense. For example, equally-sized grid cells are computationally fast but may give lower acceptance probabilities if they give coarse approximations in high-density regions. Where equally-sized grid cells give poor acceptance probabilities, for example, due to non-uniformity in high posterior regions over time, acceptance probabilities may be improved using one of the other approaches at a greater computational expense.

\section{Numerical illustrations}\label{applications}
We demonstrate the proposed PMPMH algorithm via two case studies. The first is an SSM with a simple one-dimensional Gaussian mixture state process, demonstrating how the algorithm can be implemented, and the properties of the algorithm when different practical decisions are made. We then show how a similar PMPMH implementation can be used to efficiently sample the latent states of a more challenging 2-dimensional population growth model that can display near-chaotic behavior.

In each case study, we compare the performance of the algorithm to two particle Gibbs algorithms: the particle Gibbs (PG) sampler \citep{Chopin_2015, Murphy_2016} and the particle Gibbs with ancestor sampling (PGAS) algorithm \citep{Lindsten_2014}, both with multinomial resampling. The two methods are particle MCMC algorithms and, like the proposed algorithm, sample from the joint posterior distribution of the states and parameters by updating each in turn from their full conditional distributions. The PGAS algorithm, in particular, is a state-of-the-art method that often improves upon the mixing properties of the PG algorithm by reducing sample impoverishment \citep{Nonejad_2015, Meent_2015, Kantas_2015}, though at an increased computational cost. The code used to implement the PG, PGAS, and PMPMH algorithms in each numerical illustration is given in Online Resource 1 of the Supplementary Information.

\subsection{Gaussian mixture state-space model}\label{eg:Gauss}
\begin{figure}[h]
    \centering
    \begin{subfigure}[t]{0.495\textwidth}
         \centering
    \includegraphics[width=0.4\paperwidth]{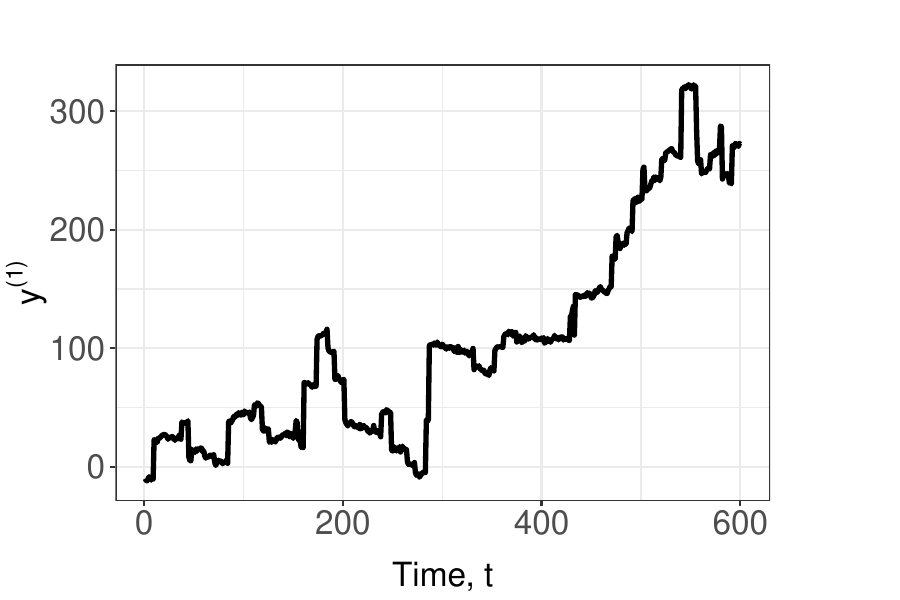}
    \caption{}
    \end{subfigure}
    \begin{subfigure}[t]{0.495\textwidth}
         \centering
     \includegraphics[width=0.4\paperwidth]{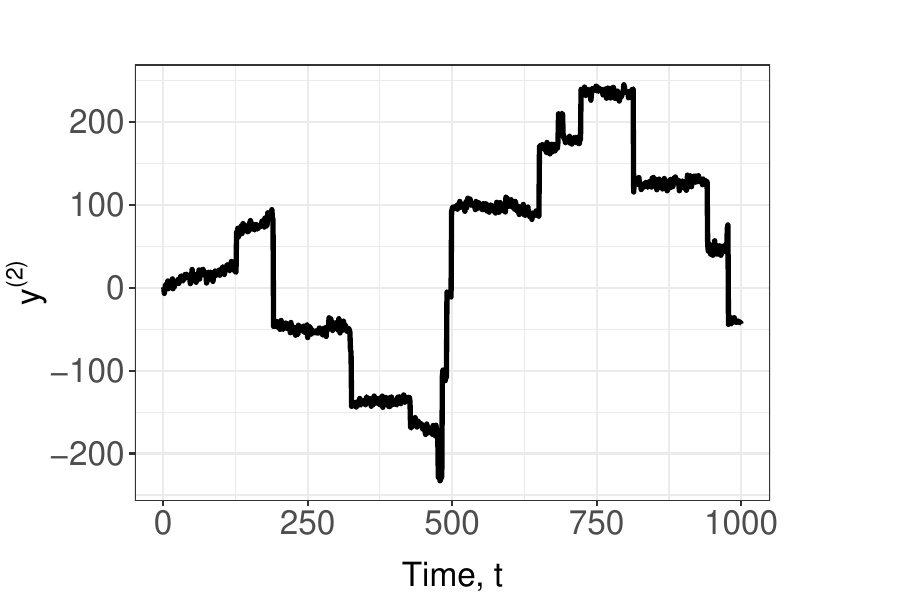}
     \caption{}
    \end{subfigure}
    \caption{Simulated data from the Gaussian mixture SSM: $y^{(1)}_{1:600}$ using (a) $\pmb{\theta}=(0.9, 1, 700, 1),$ $T=600$, and (b) $y^{(2)}_{1:1000}$ using $\pmb{\theta}=(0.99, 1, 10000, 10),$ $T=1000$}
    \label{fig:Mixnorm_data}
\end{figure}

We consider a simple one-dimensional Gaussian mixture state process:
\begin{align*}
    x_1 &\sim w_1 N(1, \sigma^2_{\eta_1}) + (1-w_1)N(1, \sigma^2_{\eta_2}),\\ 
    x_t \vert x_{t-1} &\sim w_t N(x_{t-1}, \sigma^2_{\eta_1}) + (1-w_t) N(x_{t-1}, \sigma^2_{\eta_2}), \hspace{2mm} t= 2, \dots, T,\\ 
    w_t &\sim Bernoulli(p), \hspace{2mm} t= 1, \dots, T,
\end{align*}
where $p \in [0,1]$ denotes the probability of selecting each mixture component of the state distribution (the Gaussian distributions with respective variances $\sigma^2_{\eta_1}$ or $\sigma^2_{\eta_2}$). Data, $y_{1:T}$, are observed according to $y_t \vert x_t \sim N(x_t, \sigma^2_{\epsilon})$ and the model parameters are given by $\pmb{\theta}=(p, \sigma^2_{\eta_1}, \sigma^2_{\eta_2}, \sigma^2_{\epsilon})$. We simulate two data sets from this model, $y_{1:600}^{(1)}$ using $\pmb{\theta}=(0.9, 1, 700, 1)$ and $T=600$ (Model 1), and $y_{1:1000}^{(2)}$ using $\pmb{\theta}=(0.99, 1, 10000, 10)$ and $T=1000$ (Model 2), shown in Figure \ref{fig:Mixnorm_data}.

\newpage 
Both data sets are simulated with a high-variance second mixture component, selected with a low probability, resulting in infrequent but large \emph{jumps} in the state process. The parameter values for Model 2 result in more occasional and larger jumps in the state process than the first. We compare the properties of the algorithm in both cases.

\subsubsection{PMPMH implementation}
We sample from the joint posterior distribution of the states and parameters, updating $x_{1:T}$ and $\pmb{\theta}$ from their conditional distributions. The sampling schemes and priors for updating $\pmb{\theta}$ are given in Appendix \ref{app:Mixnorm}. Here, we apply the two stages of the PMPMH algorithm to update $x_{1:T}$ conditional on $\pmb{\theta}$. The PMPMH framework presented can be adapted in several ways, for example, by changing the deterministic integration method used to approximate the HMM probabilities, the number of grid cells, and the distribution used to propose values from within grid cells. However, we show in this section that we achieve stable and efficient performance by fixing a number of these decisions and opting for the simple choices given in Appendix \ref{other_practicals}.

Instead, we focus on the practical decisions that are relevant to the efficiency of the algorithm: (a) the choice of each Approach $1-3$ of Section \ref{practical_considerations} (including the range of the state space covered by the finite grid cells, $\mathcal{S}$, which we set using scalar variances), (b) the number of grid cells, $N$, and (c) the temporal block sizes in which the states are updated, $\ell$. We test the efficiency of the algorithm under various combinations of these tuning parameters:
\begin{itemize}
    \item[(a)] the finite grid cells in Approach 1 over a range of $\mathcal{S}=150, 250, 350, 450, 550$ units for both data sets (compared to the range of each set of observations, $y_{1:600}^{(1)}$ and $y_{1:1000}^{(2)}$, equal to $334$ and $478$ units respectively). In Approaches 2 and 3, we use $\mathcal{S}=1, 3, 5, 7, 9, 11, 13, 15$ units for both models,
    \item[(b)] $N=5, 10, 20$,
    \item[(c)] $\ell=1, 4, 10$, overlapping blocks by one state to improve the mixing of the states at the ``boundaries'' of each block as suggested in \citep{Fearnhead_2011}.
\end{itemize}

\subsubsection{Results}\label{Gauss:results}

For each combination of the tuning parameters listed above, the results are based on $10$ separate runs of $10,000$ iterations each, taking $1.3-7.5$ hours with $1.6$ GHz CPU for Model 1 (depending on the number of grid cells and the approach chosen) and $2.2-11$ hours for Model 2. We present the results for each model in Tables \ref{Gauss:metrics} and \ref{Gauss:metrics2} respectively. 

\begin{table*}[ht]
    \centering
    \begin{tabular}{|c|c|c||c|c|c|c|}
    \hline \multicolumn{7}{|c|}{$\mathbf{y^{(1)}_{1:600}}$} \\
    \hline 
    \hline 
    & & \textbf{$\%$ of range} &
    \multicolumn{2}{|c|}{\textbf{Approach 2}} & \multicolumn{2}{|c|}{\textbf{Approach 3}} \\
    \textbf{N} & \textbf{$\mathbf{\mathcal{S}}$} & \textbf{of the data} & \textbf{ESS}  & \textbf{ESS/s} & \textbf{ESS} & \textbf{ESS/s}\\ 
    \hline 
    \hline 
    $5$ & $1$ & 0.3 & 100 & 0.02 & - & - \\
    & $3$ & 0.9 & 900 & 0.18 & 1000 & 0.20 \\ 
    & $5$ & 1.5 & 700 & 0.14 & 100 & 0.02 \\  
    & $7$ & 2.1 & 300 & 0.06 & - & -\\ 
    & $9$ & 2.6 & 200 & 0.04 & - & -\\ 
    & $\geq 11$ & $\geq 3.3$ & - & - & - & - \\
    
   $10$ & $1$ & 0.3 & 100 & 0.01 & - & - \\
    &  $3$ & 0.9 & 1000 & 0.10 & 1700 & 0.17  \\
   & $5$ & 1.5 & 2400 & 0.24 & 200 &  0.02 \\
    & $7$ & 2.1 & 2200 & 0.22 & - & -  \\ 
    & $9$ & 2.6 & 2000 & 0.20 & - & - \\ 
    & $\geq 11$ & $\geq 3.3$ & - & - & - & - \\

     $20$ & $1$ & 0.3 & 100 & 0.00 & - & -  \\
    & $3$ & 0.9 & 700 & 0.03 & 2200 & 0.08 \\ 
     & $5$ & 1.5 & 2700 & 0.10 & 1000 & 0.04  \\
     & $7$ & 2.1 & 3000 & 0.11 & - & - \\ 
    & $9$ & 2.6 & 3000 &  0.11 & - & - \\ 
    & $\geq 11$ & $\geq 3.3$ & - & - & - & - \\
\hline 
    \end{tabular}
    \caption{Average effective sample size (ESS) and effective sample size per second (ESS/s) for Approaches 2 and 3 of the PMPMH algorithm with temporal blocks of size $\ell=4$. The span of the finite cells is denoted by $\mathcal{S}$, the number of grid cells, $N$. Dashed lines indicate that convergence had not occurred within $10,000$ iterations. The ESS is rounded to the nearest $100$ to account for variation between chains. Since at least one model parameter was updated at each iteration for each run of the algorithm, the computational times are the same for Approaches 2 and 3: $5,000$ seconds for $N=5$, $10,000$ seconds for $N=10$, and $27,000$ seconds for $N=20$, where the computational time is rounded to the nearest $1000$ seconds to account for variations in computing speeds }
    \label{Gauss:metrics}
\end{table*}

\begin{table*}[ht]
    \centering
    \begin{tabular}{|c|c|c||c|c|c|c|}
    \hline 
        \multicolumn{7}{|c|}{$\mathbf{y^{(2)}_{1:1000}}$}  \\ 
    \hline 
    \hline
       & & \textbf{$\%$ of range} &
    \multicolumn{2}{|c|}{\textbf{Approach 2}} & \multicolumn{2}{|c|}{\textbf{Approach 3}} \\
    \textbf{N} & \textbf{$\mathbf{\mathcal{S}}$} & \textbf{of the data} & \textbf{ESS}  & \textbf{ESS/s} & \textbf{ESS} & \textbf{ESS/s}\\ 
    \hline 
    \hline 

        $5$ & $\leq 3$ & $\leq 0.6$ & - & - & - & - \\  
        & $5$ & $1$ & - & - & 200 & 0.03 \\ 
         & $7$ & $1.5$ & - & - & 500 & 0.06 \\  
         & $\geq 9$ & $\geq 1.9$ &- & - & - & -  \\  
         
         $10$ & $\leq 3$ & $\leq 0.6$ & - & - & - & - \\  
        & $5$ & $1$ & - & - & 200 & 0.02 \\ 
         & $7$ & $1.5$ &100 & 0.01 & 500 & 0.04 \\  
         & $9$ & $1.9$ & 200 & 0.02 & 200 & 0.02 \\
         & $11$ & $2.3$ & 100 & 0.01 & - & -  \\ 
          & $\geq 13$ & $\geq 2.7$ &- & - & - & - \\
          
          $20$ & $\leq 3$ & $\leq 0.6$ & - & - & - & - \\
          & $5$ & $1$ & - & - & 300 &  0.01 \\
         & $7$ & $1.5$ & 100 & 0.00  & 600 &  0.02 \\
         & 9 & 1.9 & 200 & 0.01 & 400 & 0.01 \\ 
         & 11 & 2.3 & 200 & 0.01 & 400 & 0.01 \\ 
           & $\geq 13$ & $\geq 2.7$ &- & - & - & - \\
\hline 
    \end{tabular}
    \caption{Average effective sample size (ESS) and effective sample size per second (ESS/s) for Approaches 2 and 3 of the PMPMH algorithm with temporal blocks of size $\ell=4$. The span of the finite cells is denoted by $\mathcal{S}$, the number of grid cells, $N$. Dashed lines indicate that convergence had not occurred within $10,000$ iterations. The ESS is rounded to the nearest $100$ to account for variation between chains. Since at least one model parameter was updated at each iteration for each run of the algorithm, the computational times are the same for Approaches 2 and 3: $8,000$ seconds for $N=5$, $13,000$ seconds for $N=10$, and $38,000$ seconds for $N=20$, where the computational time is rounded to the nearest $1000$ seconds to account for variations in computing speeds}
    \label{Gauss:metrics2}
\end{table*}

\newpage 
Increasing the number of grid cells, $N$, results in a more accurate approximation to $p(x_{1:T} \vert y_{1:T}, \pmb{\theta})$ in the region covered by the finite grid cells, improving mixing but at a higher computational cost. The equally-spaced grid cell Approach 1 performed poorly on both simulated models compared to the quantile-based methods of Approaches 2 and 3, requiring a large span for both models ($350$ and $550$ units respectively) due to a large range of values in the high posterior regions, and thus large numbers of grid cells and a large computational cost, to provide a reasonable HMM approximation.

Under Approaches 2 and 3, using blocks of size $\ell=10$ or a span for the finite cells greater than $11$ units required more than $N=20$ grid cells for convergence and resulted in much more costly and less efficient implementations than the other sets of tuning parameters. Further, using single-site updates ($\ell=1$) gave poor mixing when compared to blocks of size $\ell=4$ due to the correlation between consecutive states. The results when using $\ell=1, 10$, a finite span greater than $11$ units and $N>20$ are thus excluded from Tables \ref{Gauss:metrics} and \ref{Gauss:metrics2} for brevity.

The results for Approaches 2 and 3 in Table \ref{Gauss:metrics} use blocks of size $\ell=4$, and where convergent, converge within $1000-5000$ iterations using the  Brooks-Gelman-Rubin (BGR) diagnostics in \cite{Gelman_1992, Brooks_1998}. In the implementations relating to Model 1, both approaches for defining the grid cell boundaries yield similar levels of efficiency when considering the effective sample size (ESS) per second, likely since the small observation error means that both approaches focus grid cells in roughly the same region of the state space around the current state. This is potentially also the reason that both methods required a small span for the finite cells and a small number of grid cells for convergence relative to the range of the data ($1-9$ vs. $344$ units, around $0.3-2.6\%$ of the range of the data): as is true of proposal distributions that make local moves, grid cells can be focused over a smaller range of the data compared to global-move approaches and still achieve good acceptance probabilities and mixing. The smaller range also means that as few as $5$ grid cells can be used to achieve good convergence properties (compared to $50$ in Approach 1) via a good HMM approximation over the region, reducing the computational cost of the approaches. 

In comparison to Model 2, where the observation process variance is larger, Approach 2 is different from a local-move proposal distribution and generally exhibits poor mixing and convergence in comparison to Approach 3. Conversely, the local moves of Approach 3 are effective at achieving convergence and exhibit relatively stable performance even for $N=5$. The efficient ranges for the finite grid cells are similar to those in the implementations of the first model as a percentage of the range of the data (here, $1-2.3\%$ of the range of the data).

We also fitted both of the models using the PG and PGAS algorithms with $5$ to $1000$ particles, and various combinations of resampling thresholds based on the standard percentage ESS criterion \citep{Cappe_2005}. The PG sampler did not converge for either model using as many as $1000$ particles, resulting in computational times of around 14 hours for Model 1 and 28 hours for Model 2 on $1.6$ GHz CPU. Conversely, the PGAS sampler converged using as few as $5$ particles for both models. On the whole, the PGAS sampler gave greater levels of efficiency than the PMPMH algorithm for both of the models, achieving an average ESS per second of around $4.45$ for Model 1 and $1.49$ for Model 2. However, this efficiency was not uniform across all states. The average ESS per second of the states simulated according to the second mixture was $0.25$ to $0.41$ for Model 1 and $0.004$ to $0.008$ for Model 2. In contrast, where convergent, the PMPMH algorithm was more robust to the mixture associated with the state, with the ESS per second of states in the second mixture minimally $99\%$ of those quoted in Tables \ref{Gauss:metrics} and \ref{Gauss:metrics2} (the average ESS per second in the second mixture-distributed states ranging from $0.01$ to $0.24$ for Model 1 and $0.01$ to $0.06$ for Model 2). We now investigate the comparative performance of the algorithm on a model that can display near-chaotic behavior.

\subsection{Nicholson's blowfly model}\label{eg:Blowflies}
We consider Nicholson's Blowfly model for chaotic population growth in \cite{Wood_2010}. The population counts over time, denoted by $N_{1:T}$, arise from two correlated survival and birth processes, denoted $S_{1:T}$ and $R_{\tau+1:T}$, $\tau>0$, respectively. Following \cite{Wood_2010}, we let $\exp(-\delta \epsilon_t)$ denote the daily survival probability with associated environmental error term $\epsilon_t$, such that $\epsilon_t \sim \Gamma(\beta_{\epsilon}, \beta_{\epsilon})$, $\beta_{\epsilon}>0$. The survival component of the system process is given for $t=1, \dots, T$ by

\begin{equation*}
    S_t \sim Binom(N_{t-1}, \exp(-\delta \epsilon_t)),
\end{equation*}

\begin{figure}[H]
    \centering
    \includegraphics[width=0.6\paperwidth]{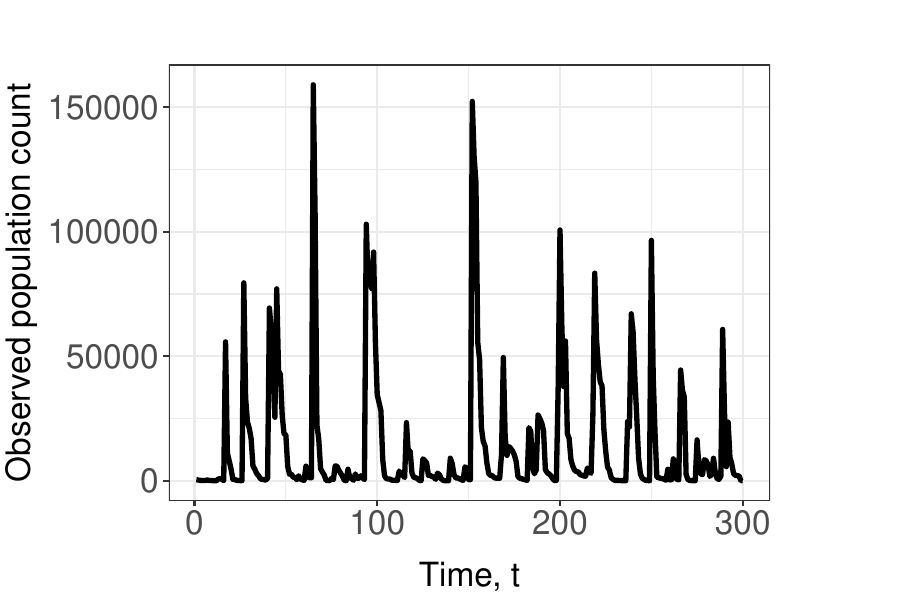}
    \caption{Simulated blowfly population count data using $T=300$ and $\pmb{\theta}=(0.7, 50, 1, 0.1, 1)$}
    \label{fig:Nich_data}
\end{figure}

Letting $e_t$ denote an environmental noise term in the reproductive process such that $e_t \sim \Gamma(\beta_e, \beta_e)$, $\beta_e>0$, the reproductive component of the system process is given for $t=\tau+1, \dots, T,$ by 

\begin{equation*} 
   R_t \sim Po\left(PN_{t-\tau-1} \exp(-\frac{N_{t-\tau-1}}{N_0}) e_t\right),
\end{equation*}

where $N_t = S_t + R_t$ for $t=\tau +1, \dots, T$ and $N_t=S_t$ for $t=1, \dots, \tau$. We let $\tau=5$ be the known time lag between population count and subsequent birth count, and $N_0=50$ is the known initial population count. Further, $\epsilon_{1:T}$ and $e_{\tau+1:T}$ are known. The survival and birth processes, $S_{1:T}$ and $R_{\tau+1:T}$, are unknown latent states, with observed population counts $y_t \sim Po(\phi N_t)$, for all $t$. Further, the model parameters $\pmb{\theta}=(\delta, P, \beta_\epsilon, \beta_e, \phi)$ are unknown and positively valued. Figure \ref{fig:Nich_data} shows the simulated data used in this case study, $y_{1:T}$, using $\pmb{\theta}=(0.7, 50, 1, 0.1, 1)$ and $T=300$.

\subsubsection{PMPMH implementation}
We devise a Metropolis-Hastings scheme targeting the joint posterior distribution of the unknown latent states and model parameters, $p(S_{1:T}, R_{\tau+1:T}, \pmb{\theta} \vert y_{1:T})$. The model parameters are updated from their full conditional distribution, $p(\pmb{\theta} \vert S_{1:T}, R_{\tau+1:T}, y_{1:T})$, using the priors and sampling schemes given in Appendix \ref{app:Blowflies}. In this section, we describe how the PMPMH algorithm can efficiently update the latent states, $S_{1:T}$ and $R_{\tau+1:T}$, conditional on $\pmb{\theta}$.

First, the state process is now two-dimensional at each time point. To simplify the design of the grid used within the PMPMH algorithm, we build separate PMPMH proposal distributions targeting the full conditional distributions in each state dimension, $p(S_{1:T} \vert R_{\tau+1:T}, y_{1:T}, \pmb{\theta})$ and $p(R_{\tau+1:T} \vert S_{t:T}, y_{1:T}, \pmb{\theta})$, respectively. To sample from each of the full conditional distributions in turn, a number of the practical decisions remain the same as in the previous example of Section \ref{eg:Gauss}. We first fix several of the practical decisions to those in Appendix \ref{other_practicals} and use blocks of size $\ell=4$, overlapping blocks by one state to reduce the correlation between states at the block boundaries. We test the performance of the algorithm for $N \in \{10, 20, 50\}$ grid cells.

There are two main differences in this implementation compared to the previous example. The first accounts for the near-chaotic state processes and the large range of the data ($1.59 \times 10^{5}$ units). We apply the current state-centered Approach 3 of Section \ref{practical_considerations}, which requires fewer grid cells for good mixing properties due to the lower span for the finite cells required to achieve efficient (more local) moves. Within this approach, we permit large variability around large values for the state process by adjusting the variance of the Gaussian distributions used to set the boundaries of the grid cells: we set the variance proportional to the current state at each time point. We try factors of proportionality of $0.1, 0.25, 0.5, 1$, with finite grid cells between the $q$ and $1-q$ quantiles with $q=0.01, 0.1, 0.2$, resulting in (average) spans for the finite cells ranging from $\mathcal{S}=52-400$. The second difference is that we ensure the grid cells are over a discrete space bounded at zero by rounding the quantiles used to determine the grid cells, setting the lower boundary to zero if needed.

\subsubsection{Results}
We assessed the performance of the PMPMH algorithm using $10$ independent MCMC simulations of $50,000$ iterations and present the performance metrics in Table \ref{tab:Nich}. Where the implementations of the PMPMH algorithm converged, convergence was achieved within $4,000-18,000$ iterations, taking $11-27$ hours using $1.6$ GHz CPU. Fairly consistent and efficient performance (in terms of ESS per second, Table \ref{tab:Nich}) is achieved using $20-50$ grid cells with the finite cells covering a region of the state space $0.05-0.25\%$ of the range of the data ($\mathcal{S} \in [82, 400]$).

 \begin{table*}[h]
    \centering
    \begin{tabular}{|c|c|c|c|c|}
    \hline 
        \textbf{Max N} &  \multicolumn{2}{|c|}{\textbf{Average S}} & &  \\
 &  \textbf{$\mathbf{\mathcal{S}}$} & \textbf{R} & \textbf{ESS}  & \textbf{ESS/s} \\ 
        \hline 
        10 &$\leq 76$ & $\leq 70$ & - & - \\
        
        & 82-100 & 75-90 & 2500 & 0.05 \\ 
        
        & 100-200 & 90-180 & 3200 & 0.07 \\ 
        
        & $>200$ & $>180$ & - & - \\

        20 & $\leq 76$ & $\leq 70$ & - & - \\ 
        & 82-100 & 75-90 & 4400 & 0.09  \\ 
        & 100-200 & 90-180 & 5400 & 0.09 \\ 
        & 200-300 & 180-260 & 7000 & 0.11 \\ 
        & 300-400 & 260-300 & 6700 & 0.10\\

        50 &  $\leq 76$ & $\leq 70$ & - & - \\
        & 82-100 & 75-90 & 4400 & 0.06 \\ 
        
        & 100-200 & 90-180 & 5700 & 0.06 \\ 
        
        & 200-300 & 180-260 & 7300 & 0.08 \\ 
        
        & 300-400 & 260-300 &  6900 & 0.07 \\ 
            \hline 
    \end{tabular}
    \caption{Average ESS and ESS/s across the state and parameter samples for various sets of tuning parameters under Approach 3 with blocks of size $\ell=4$. Since the span of the finite cells varies across time points, we provide the average span provided by the MCMC output. The ESS has been rounded to the nearest 100 to account for the variability within the chains}
    \label{tab:Nich}
\end{table*} 

Approach 3 mixes effectively by producing a reasonable HMM approximation in a relatively small region around the current state and is also computationally cheap to implement. The efficiency of the algorithm reduces at an upper bound of $50$ grid cells, indicating that the improved mixing properties of the algorithm are not justified by the extra cost when compared with using fewer grid cells. Conversely, $10$ grid cells, although lower in computational cost, do not produce a sufficiently accurate approximation to the posterior in the region of the current state. We also note that, similarly to the previous case study, implementations relying on \emph{extremely} small moves around the state space (finite cells $<0.05\%$ of the range of the data) gave poor mixing and convergence properties. 
 
The PGAS sampler, when applied to the model using both joint updating of the latent states and updating each latent state process in turn from their full conditional distributions, did not converge in $50,000$ iterations with $1000$ particles, remaining in a range of the state space unrepresentative of the posterior due to sample impoverishment and taking $160$ hours with $1.6$ GHz CPU. 

\section{Discussion}\label{discussion}
We provide an efficient novel approach for fitting general SSMs to observed data. The idea uses tractable HMM approximations to efficiently update the unobserved latent states in a Metropolis-Hastings algorithm. We demonstrate the generality of the proposed approach by its application to two problems, including a challenging near-chaotic problem. The proposed PMPMH algorithm is demonstrated to provide reliable posterior estimates within reasonable computational time frames, especially when compared to 
a state-of-the-art method, which did not converge within a reasonable time frame for the near-chaotic problem. The flexibility of the PMPMH approach via the tuning parameters, including the location, size, and number of grid cells, and the temporal block size provides an adaptable and efficient algorithm. Using simple methods for placing the grid cells, such as equally-sized grid cells, can be efficient since they are computationally cheap. However, such simple methods may require a large number of grid cells to achieve good mixing if the range of the data is large. A current state-centered approach for approximating the HMM is useful, especially when the state space covers a large range, and using a data-centered approach is useful provided the range of the data is small. 

The PMPMH approach motivates several interesting points for future research. One avenue for research is to consider the computational efficiency of parallelization when applied to different components of the algorithm. The computational efficiency of algorithms using block-updating strategies can be improved via parallelization \citep{King_2023}. Within the framework of the PMPMH algorithm specifically, serially independent blocks of states can be updated in parallel within each Metropolis-Hastings iteration. Further, for the state-centered grid cell approach presented here, computing the approximate HMM probabilities in each block in parallel can lead to an improvement in the execution time of the approximate HMM computations. However, as with parallelization schemes for particle MCMC methods, such as in \cite{Henriksen_2012}, the user must balance memory limitations and the computational cost from re-synchronization with the computational gains from parallel implementation. Within the sequential computing context of this paper, one way to reduce the overall computational cost of the algorithm is to reduce the number of times the HMM approximation is calculated across iterations. This is valid for approximation methods that do not depend on the current state or model parameters. However, for deterministic grids that adapt to previous samples (for example, Approaches 2 and 3), the resulting posterior estimate may become biased since the chain is no longer Markovian \citep{Haario_1999}. However, \cite{Haario_1999} also show that the bias introduced is negligible in some applications and that unbiased samples can be obtained by introducing some pre-determined stopping criterion for updating the grid, that is, specifying a number of iterations beyond which the grid is no longer updated. The computational gains from updating the HMM approximation less frequently could, however, be balanced with potentially reduced mixing properties. For example, local-move grids using a low span for finite cells are more sensitive to the frequency of the HMM approximation since they rely on proposed moves being made in the region of the current state. However, updating the grid less frequently may give an efficient approach when applied to more global-move samplers.

One may also attempt to find efficient ways to adapt the HMM approximations. With any such adaptations, there is a trade-off between the improved mixing properties and the computational cost arising from the complexity of the method. For example, in Step 1 of the algorithm, more complex numerical integration methods could be used to approximate the SSM density in each grid cell. Such methods include increasing the number of points in each grid cell used for approximation calculations, simple linear functions joining evaluations at midpoints, or other fast evaluations that include the gradient in each grid cell. There may also be ways to improve the efficiency of the within-cell proposal distributions in Step 2, including approaches that approximate the posterior density in each grid cell where the conditional posteriors are irregular or vary in an unsystematic way depending on the value of the parameter. These approaches could be valuable to explore for more complex SSMs, in which the improved mixing may justify the added computational cost. Further, these approaches to improving mixing in each step may permit the use of fewer grid cells, decreasing the overall computational cost, which may also permit greater scalability of the HMM to higher dimensional latent spaces. Although we demonstrated the scalability of the proposed algorithm to higher dimensional spaces by updating each state dimension conditionally on the remaining state dimensions, higher dimensional spaces are a particular challenge if individual state dimensions are highly correlated, resulting in poor mixing. In this case, if low-dimensional sets of state dimensions are independent, for example, Factorial SSMs \citep{Ghahramani_1995, Rimella_2022}, or if dimensions of the state space are partially integrable \citep{Borowska23}, the poor mixing from highly-correlated state dimensions could be improved by performing joint updates using lower-dimensional transition matrix approximations. However, in general, the issue of the scalability of grid-based methods to high-dimensional SSMs is a challenge to the proposed algorithm and an active area of research. 

\section{Supplementary information}
\noindent \textbf{Online Resource 1} \quad GitHub repository containing the R code for the implementations of the PMPMH, PG, and PGAS algorithms in Section \ref{applications}: \newline \url{https://github.com/mallewellyn/PMPMH}.

\section*{Declarations}

\noindent \textbf{Funding} M.L. was supported by studentship funding from the Engineering and Physical Sciences Research Council (EPSRC). R.K. was supported by the Leverhulme Research Fellowship (RF-2019-299). V. E. was supported by the \emph{Agence Nationale de la Recherche} of France under PISCES (ANR-17-CE40-0031-01), the Leverhulme Research Fellowship (RF-2021-593), and by ARL/ARO (W911NF-22-1-0235).\\

\noindent \textbf{Conflcts of interest} The authors report that there are no conflicts of interest to declare.\\

\noindent \textbf{Code availability} See Supplementary Information.

\begin{appendices}

\section{Parameter prior distributions}\label{app:2}

\subsection{Gaussian Mixture State-Space Model Parameters in \linebreak Section \ref{eg:Gauss}}\label{app:Mixnorm}
In Section \ref{eg:Gauss}, we conditionally update the parameters from both $p(\pmb{\theta} \vert x_{1:T}, y_{1:T}^{(1)})$ and $p(\pmb{\theta} \vert x_{1:T}, y_{1:T}^{(2)})$ using single-site conjugacy in the observation variance parameter. For both Model 1 and Model 2, we assign vague priors to enable comparison with the simulated parameter for diagnostic purposes. For Model 1,
\begin{equation*}
    \sigma_{\epsilon}^2 \sim \Gamma^{-1}(2, 2),
\end{equation*}
resulting in a Gibbs sampler for $ \sigma_{\epsilon}^2$. We assign further high-variance priors to the other parameters in Model 1:
\begin{align*}
p &\sim U(0,1),\\
\sigma_{\eta_1}^2 &\sim \Gamma^{-1}(2,2),\\
\sigma_{\eta_2}^2 &\sim \Gamma^{-1}(2,700),
\end{align*} 
where $p$, $\sigma_{\eta_1}^2$ and $\sigma_{\eta_2}^2$ are proposed at iteration $m$ from uniformly-distributed random walk proposals over intervals of length $0.3$, $2$, and $160$ units respectively. For Model 2, we similarly assign the independent priors: 
\begin{align*}
\sigma_{\epsilon}^2 &\sim \Gamma^{-1}(2, 10),\\
p &\sim U(0,1),\\
\sigma_{\eta_1}^2 &\sim \Gamma^{-1}(2,2),\\
\sigma_{\eta_2}^2 &\sim \Gamma^{-1}(2,700),
\end{align*} 

where the prior for $\sigma_{\epsilon}^2$ again results in Gibbs steps for this parameter. We propose $p$, $\sigma_{\eta_1}^2$ and $\sigma_{\eta_2}^2$ 
at iteration $m$ from a uniform random walk proposal distribution over $0.02$, $0.5$, and $20000$ units respectively. All intervals for the random walk proposal distributions are set using pilot tuning.

\subsection{Nicholson's Blowfly Model Parameters in Section \ref{eg:Blowflies}}\label{app:Blowflies}
We give details for the priors and sampling schemes for the model parameters, $\pmb{\theta}$, in \linebreak Section \ref{eg:Blowflies}. Once again, we make use of the conjugate priors for single site updates of $\pmb{\theta}$ where possible and assign vague priors to help diagnose convergence:
\begin{align*}
    \delta &\sim \Gamma(0.007, 0.01),\\
    P &\sim \Gamma(50, 1),\\
    \beta_{\epsilon} &\sim \Gamma^{-1}(100, 100),\\
    \beta_e &\sim \Gamma^{-1}(10, 1),\\
    \phi &\sim \Gamma(0.01, 0.01).
\end{align*}
This results in single-site Gibbs updates for $P$ and $\phi$. For $\delta$, $\beta_{\epsilon}$ and $\beta_e$ and we use a random walk Metropolis-Hastings step with uniform proposal distributions over intervals of length $0.03$, $0.5$, and $0.05$ respectively. The parameters of the proposal distributions were chosen via pilot tuning.

\section{Practical decisions}\label{other_practicals}
We make a number of practical decisions in order to implement the PMPMH algorithm in our case studies. We opt for simple choices for most of these and discuss these here.

\subsection{Deterministic integration method}\label{param:1}
In  Section \ref{method:grid_cell_traj}, we focus on the simple case for approximating the HMM probabilities in each grid cell: midpoint integration using $S=1$. However, the Riemann sum integral approximation method scales simply for higher order polynomials where $S\geq 2$, or these methods can be easily replaced by more complex numerical integration strategies. However, the complexity of these methods should be balanced with their associated computational cost to ensure efficiency in this step. We therefore test the efficiency of the algorithm using midpoint integration ($S=1$)

The \emph{midpoint} and length of the cells must be defined to apply midpoint integration, even when the $n=1$ or $n=N$ grid cells have an infinite range. In these infinite cells, we simply set this artificial length at the average length of the finite cells, and we set the artificial midpoint at half that length away from the corresponding upper or lower boundary of the finite cells. Further, to ensure that the HMM probabilities are sufficiently high to avoid the proposal distribution resulting in a `near-reducible' Markov chain, we set a lower bound on the HMM probabilities. In all the implementations of this paper, we use a lower bound of $0.01$ and renormalize so that they sum to one. 
 
 \subsection{Proposal distributions within the grid cells}\label{param:2}
 Once we have sampled a set of grid cells, indexed by $b'_{1:T}$ as per Section \ref{method:point_traj}, we sample point values from within the grid cells using simple proposal distributions for each $t\geq 1$. In all implementations, we sample from uniform distributions with domain in the finite grid cells. 
 
 To sample values for the state in infinite grid cells in the first case study of Section \ref{eg:Gauss}, we sample from a truncated Gaussian distribution with mean equal to the finite boundary and a variance of $5$. The variance of the infinite-cell distribution is relatively low so that proposals in this grid cell are mostly concentrated around the boundary of the finite cells, increasing the density in the tails of the proposal distribution. In the second case study, we sample from a Poisson distribution with mean parameter equal to $2$ (again to ensure a relatively heavy-tailed proposal distribution), shifted to the lower bound of the upper (infinite) grid cell.
 
\section{Pseudocode for the PMPMH algorithm with block updates}\label{code:blocking}
Here we present the pseudocode describing the two stages of the PMPMH algorithm with the states updated in blocks. This supplements the description of the block-updating procedure in Section \ref{method:blocking}.  

Note that, in this case, the acceptance probability for a proposed set of $\ell$ states starting at time $t$, $x'_{t:t+\ell-1}$, is given by

\begin{align}
    &p(x_{t:t+\ell-1}, x'_{t:t+\ell-1} \vert \pmb{\theta}) = \nonumber \\ 
    &\quad \frac{p(x'_{t:t+\ell-1} \vert x_{t-1}, x_{t+\ell}, y_{t:t+\ell-1}, \pmb{\theta})}{p(x_{t:t+\ell-1} \vert x_{t-1}, x_{t+\ell}, y_{t:t+\ell-1}, \pmb{\theta})} \times \nonumber \\
    &\quad \quad \frac{ q(x_{t:t+\ell-1} \vert x_{t-1}, x_{t+\ell}, y_{t:t+\ell-1}, \pmb{\theta})}{q(x'_{t:t+\ell-1} \vert x_{t-1}, x_{t+\ell}, y_{t:t+\ell-1}, \pmb{\theta})} \label{eq:blocking}
\end{align}

where, in the first block, 
\begin{flalign*}
    p(x_{t:t+\ell-1} &\vert x_{t-1}, x_{t+\ell}, y_{t:t+\ell-1}, \pmb{\theta}) =\\
    &\quad p(x_{t:t+\ell-1} \vert x_{t+\ell}, y_{t:t+\ell-1}, \pmb{\theta}),
\end{flalign*}

and similarly for the proposal density. Likewise, in the last block,
\begin{flalign*}
    p(x_{t:t+\ell-1} &\vert x_{t-1}, x_{t+\ell}, y_{t:t+\ell-1}, \pmb{\theta}) =\\
    &\quad p(x_{t:t+\ell-1} \vert x_{t-1}, y_{t:t+\ell-1}, \pmb{\theta}).
\end{flalign*}

\pagebreak 
\newpage 

\begin{algorithm*}[!h]
\caption{PMPMH algorithm: block updates}\label{PMPMH_alg_block}
\textbf{Input:} initial values $x_{1:T}^{(0)}$, $\pmb{\theta}^{(0)}$, a number of iterations $M$. A blocking strategy for the states: a block size $\ell$ and a set of starting points for each block $\{1, \dots, D\} \subset \{1, \dots, T\}$. A Gibbs or Metropolis-Hastings sampling scheme to update $\pmb{\theta}$ from $p(\pmb{\theta} \vert x_{1:T}, y_{1:T})$. 

\hrulefill

For $1 \leq t \leq \ell$ (block $d=1$), define
\vspace{2mm}
\begin{center}
$\hat{P}(B_t=k \vert B_{d-1}=b_{d-1}, y_{d:t}, \pmb{\theta})=\hat{P}(B_t=k \vert y_{d:t}, \pmb{\theta}),$
\end{center}

\begin{center}
$\hat{P}(B_t=k \vert B_{d-1}=b_{d-1}, B_{t+1}=b_{t+1}, y_{d:d+\ell-1}, \pmb{\theta})=\hat{P}(B_t=k \vert B_{t+1}=b_{t+1}, y_{d:d+\ell-1}, \pmb{\theta}).$
\end{center}

\vspace{2mm}
Also, for $t=T$ in block $D$ define
\vspace{2mm}
\begin{center}
$\hat{P}(B_t=k \vert B_{d-1}=b_{d-1}, B_{t+1}=b_{t+1}, y_{d:t}, \pmb{\theta})=\hat{P}(B_T=k \vert B_{d-1}=b_{d-1}, y_{d:t}, \pmb{\theta}).$
\end{center}

\hrulefill

\textbf{for} iterations $m=1, \dots M$ \textbf{do}

\begin{algorithmic}[1]
\State{Update $\pmb{\theta}^{(m)}$ from $p( \pmb{\theta} \vert x_{1:T}^{m-1}, y_{1:T})$}
\Statex 
\For{$d=1, \dots, D$}
\State{Define intervals with indices $B_{\max(1, d-1):\min(d+\ell, T)}$}
\For{$t=d, \dots d+\ell-1$} \Comment{(Step 1)}
\For{$k=1, \dots N$}
\State{calculate $\hat{P}(B_t=k \vert B_{d-1}=b_{d-1}, y_{d:t}, \pmb{\theta}^{(m)})$}
\EndFor
\EndFor

\Statex 
\State{sample $b'_{d+\ell-1}$ from}
\Statex{\hspace{18mm}$\{k, \hat{P}(B_{d+\ell-1}=k \vert B_{d+\ell}=b_{d+\ell}, B_{d-1}=b_{d-1}, y_{d:d+\ell-1}, \pmb{\theta}^{(m)})\}_{k=1}^N$}

\For{$t=d+\ell-2, \dots, d$}
\State{sample $b'_t$ from}
\Statex{\hspace{25mm}$\{k, \hat{P}(B_t=k \vert  B_{t+1}=b'_{t+1}, B_{d-1}=b_{d-1}, y_{d:t}, \pmb{\theta}^{(m)})\}_{k=1}^N$}
\EndFor 

\Statex 
\For{$t=d, \dots, d+\ell-1$}\Comment{(Step 2)}
\State{sample $x'_t$ from  $q(x_t \vert B_t=b'_t)$} 
\EndFor

\State{sample $u \sim U(0,1)$}

\If{$u<p(x_{d:d+\ell-1}^{(m-1)}, x'_{d:d+\ell-1} \vert \pmb{\theta})$} \Comment{Equation (\ref{eq:blocking})}
\State{$x_{d:d+\ell-1}^{(m)} \leftarrow x'_{d:d+\ell-1}$}
\Else{
$x_{d:d+\ell-1}^{(m)} \leftarrow x^{(i-1)}_{d:d+\ell-1}$} \label{acc_end_block}
\EndIf 

\EndFor

\end{algorithmic}
\end{algorithm*}

\end{appendices}

\newpage


\end{document}